\documentclass[11pt]{article}
\usepackage{fleqn,cospar}

\usepackage{url}

\usepackage{graphicx}
\usepackage[figuresright]{rotating}

\def\aox{$\alpha_{\rm ox}$}

\def\simgt{\lower 2pt \hbox{$\, \buildrel {\scriptstyle >}\over {\scriptstyle \sim}\,$}}
\def\simlt{\lower 2pt \hbox{$\, \buildrel {\scriptstyle <}\over {\scriptstyle \sim}\,$}}

\def\asca{{\it ASCA\/}}
\def\chandra{{\it Chandra\/}}
\def\conx{{\it Constellation-X\/}}

\def\genx{{\it Generation-X\/}}
\def\hst{{\it {\it HST}\/}}

\def\rosat{{\it ROSAT\/}}
\def\xeus{{\it XEUS\/}}
\def\xmm{{\it XMM-Newton\/}}

\title{X-RAYS FROM THE FIRST MASSIVE BLACK HOLES}

\author{W.N.~Brandt,\address{Department of Astronomy and Astrophysics, 
525 Davey Laboratory, The Pennsylvania State University, University Park, 
PA 16802, USA} 
C.~Vignali,$^{1}$ 
D.P.~Schneider,$^{1}$ 
D.M.~Alexander,$^{1}$ 
S.F.~Anderson,\address{University of Washington, Department of Astronomy, Box 351580, Seattle, WA 98195, USA}  
F.E.~Bauer,$^{1}$ 
X.~Fan,\address{Steward Observatory, The University of Arizona, 933 North Cherry Avenue, Tucson, AZ 85721, USA}
G.P.~Garmire,$^{1}$ 
S.~Kaspi,\address{School of Physics and Astronomy and the Wise Observatory, The Raymond and Beverly Sackler
Faculty of Exact Sciences, Tel-Aviv University, Tel-Aviv 69978, Israel} and
G.T.~Richards$^{1,}$\address{Princeton University Observatory, Princeton, NJ 08544, USA}} 

\begin{document}

\maketitle


\begin{abstract}
X-ray studies of high-redshift ($z>4$) active galaxies have advanced
substantially over the past few years, largely due to results from the
new generation of X-ray observatories. As of this writing X-ray emission 
has been detected from nearly 60 high-redshift active galaxies. This paper
reviews the observational results and their implications for models
of the first massive black holes, and it discusses future prospects for 
the field.
%
%
\end{abstract}


\section*{INTRODUCTION AND IMPORTANCE OF HIGH-REDSHIFT X-RAY STUDIES}

\

Understanding of the X-ray emission from $z>4$ active galactic 
nuclei (AGN) has advanced rapidly over the past few years. The high 
sensitivities of \chandra\ and \xmm\ have allowed the efficient detection 
of many $z>4$ AGN, and wide-field AGN surveys (e.g., the 
Sloan Digital Sky Survey, hereafter SDSS; York et~al. 2000) have 
greatly enlarged the number of suitable X-ray targets. The number 
of X-ray detections at $z>4$ has increased 
to 57 (see Figure~1a).\footnote{See 
http://www.astro.psu.edu/users/niel/papers/highz-xray-detected.dat 
for a regularly updated list of $z>4$ X-ray detections.} 
This increase has allowed the first reliable X-ray population studies 
of $z>4$ AGN, and it has been possible to obtain X-ray detections
at redshifts up to $z=6.28$ (e.g., Brandt et~al. 2002a).  

The X-ray emission from AGN provides direct information about their 
black hole regions, where accretion and black hole growth occur; the 
emission is thought to be produced by the inner accretion 
disk and its corona (e.g., Poutanen 1998). {\it Do the early black holes 
seen at high redshift feed and grow in the same way as local ones?\/} X-ray 
observations of high-redshift AGN can address this fundamental 
question. It is plausible that high-redshift AGN could be feeding 
and growing differently. The comoving number density of luminous quasars
changes by a factor of $\simgt 100$ over the history of the 
Universe (e.g., Figure~9 of Fan et~al. 2001 and references therein), 
and part of this strong evolution is believed to 
be due to environmental changes that might also impact the X-ray 
emission region. There have also been theoretical predictions 
that AGN accretion rates, relative to the Eddington rate, should 
change with redshift (e.g., Kauffmann and Haehnelt 2000); such changes can lead to accretion-disk 
instabilities (e.g., Gammie 1998 and references therein) and 
radiation ``trapping'' effects (e.g., Begelman 1979). 

X-ray absorption measurements can also be used to probe the large-scale
environments of high-redshift AGN. Changes in the amount of X-ray 
absorption with redshift have been discussed by many authors.
The fraction of radio-loud quasars (RLQs) with heavy X-ray absorption appears to 
rise with redshift, with column densities of $\simgt 2\times 10^{22}$~cm$^{-2}$ 
being seen at $z\simgt 3$
(e.g., Elvis et~al. 1998; Fiore et~al. 1998; Reeves and Turner 2000). The 
absorbing gas may be circumnuclear, located in the 
host galaxy, or entrained by the radio jets. 
Radio-quiet quasars (RQQs) definitely show less of an absorption increase with 
redshift than do RLQs (e.g., Fiore et~al.\ 1998). However, as discussed below, 
the X-ray absorption properties of high-redshift RQQs are only now becoming clear.

\begin{figure}
\includegraphics[width=90mm]{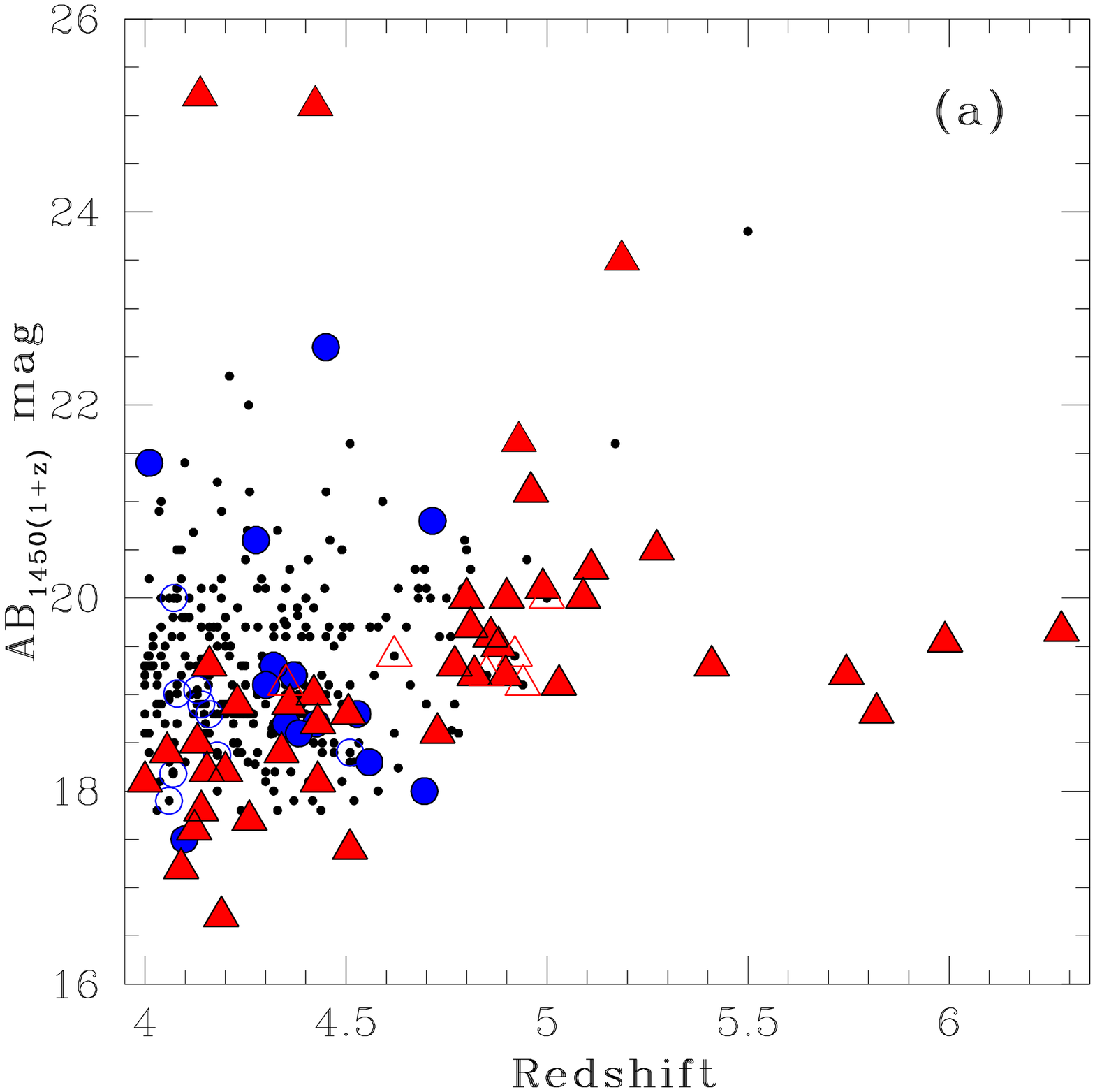}
\includegraphics[width=90mm]{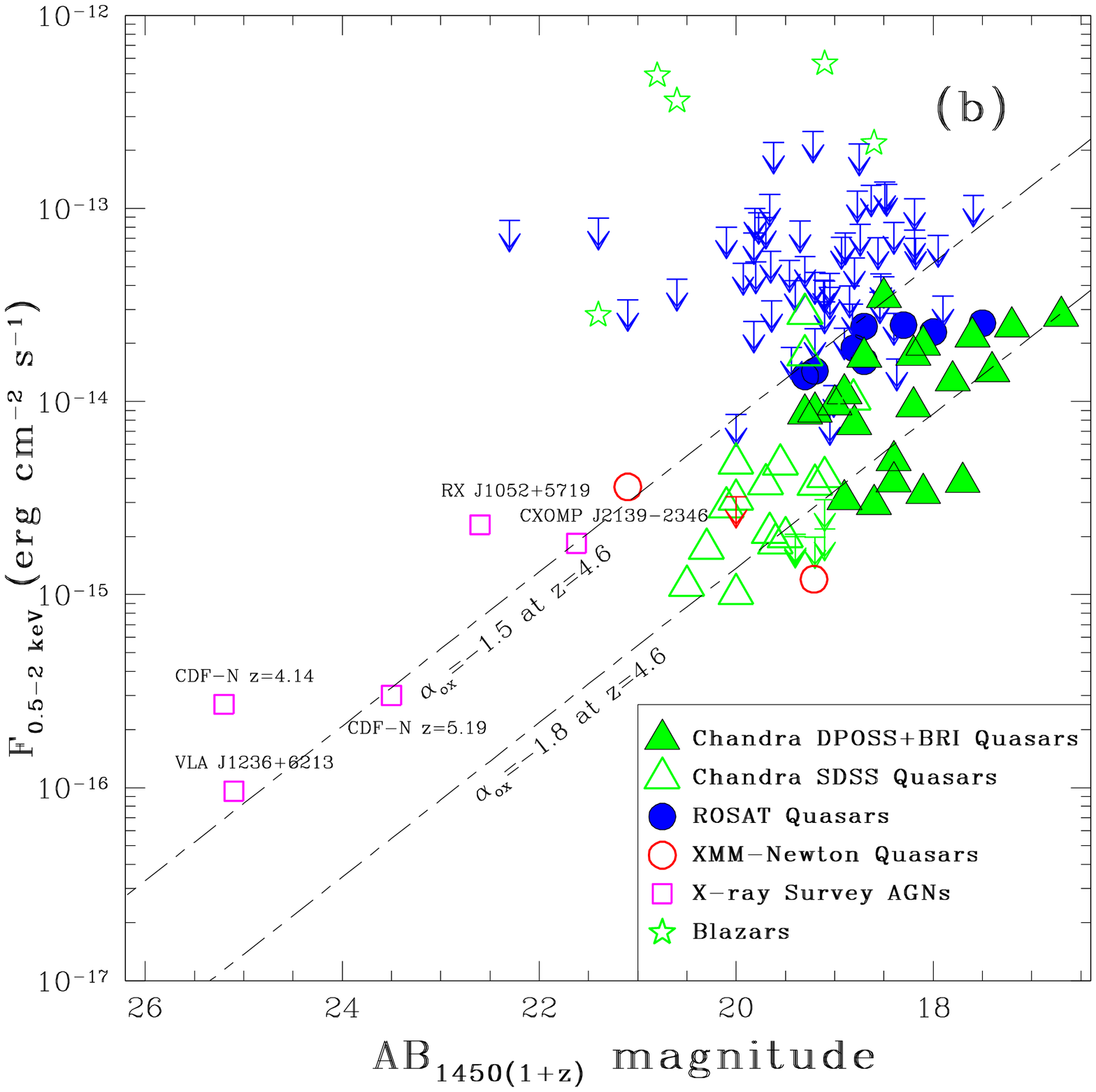}
\vspace*{-0.4 in}
\caption{(a) $AB_{1450(1+z)}$ magnitude versus redshift for $z>4$ AGN
(small solid circles). 
Solid triangles (large solid circles) indicate AGN that have been 
detected by \chandra\ or \xmm\ (\rosat).
Open triangles (large open circles) indicate AGN that have tight 
X-ray upper limits from \chandra\ or \xmm\ (\rosat). 
(b) Observed-frame, Galactic absorption-corrected \hbox{0.5--2~keV} flux 
versus $AB_{1450(1+z)}$ magnitude for $z>4$ AGNs. Object types are
shown in the key; see Figure~4 of Vignali et~al. (2003a) for further
explanation. The slanted lines show \aox=$-1.5$ and \aox=$-1.8$ loci 
at a fiducial redshift of $z=4.6$.}
\end{figure}

We adopt $H_0=65$~km~s$^{-1}$ Mpc$^{-1}$, 
$\Omega_{\rm M}=1/3$, and $\Omega_{\Lambda}=2/3$
throughout. This review partially 
updates Brandt et~al. (2002b); some of the technical details 
and references in that review are not repeated here. 


\section*{RECENT X-RAY OBSERVATIONS AND RESULTS}

\vspace*{0.4 cm}

\subsection*{X-ray Observations}

Most $z>4$ X-ray detections to date have been obtained via moderate-depth
[$\approx$~(1--20)$\times 10^{-15}$~erg~cm$^{-2}$~s$^{-1}$ in the 0.5--2~keV
band] follow-up observations of AGN discovered at other wavelengths
(e.g., Kaspi et~al. 2000; Vignali et~al. 2001; 
Brandt et~al. 2002a; Bechtold et~al. 2003; 
Vignali et~al. 2003a, 2003c).\footnote{In a handful of cases, 
moderate-depth X-ray data were serendipitously available for a field in 
which a $z>4$ quasar was subsequently discovered (predominantly using
data at other wavelengths).} Due to its superb 
imaging capability, \chandra\ is particularly 
effective at such observations; a measurement of only
$\approx$~2--3 photons at the known position of a quasar constitutes
a significant \chandra\ detection! \chandra\ ``snapshot''
observations of only 4--10~ks are now reliably yielding X-ray detections
of many $z\approx$~4--6 quasars. 
For example, we have made 30 such snapshots of $z>4$ quasars and have
detected 87\% of our targets; we have 25 more snapshots scheduled for \chandra\ 
Cycle~4. Our targets have included the highest redshift SDSS quasars 
($z\simgt 4.8$; $M_{\rm B}\approx -27.0$ to $-28.5$) 
and 
the most luminous Palomar Digital Sky Survey (DPOSS; Djorgovski et~al. 1998) quasars 
($z\approx$~4.1--4.5; $M_{\rm B}\approx -28.5$ to $-30.5$).
We are also now targeting RLQs and ``exotic'' quasars 
(e.g., the quasars lacking emission lines in Fan et~al. 1999 
and Anderson et~al. 2001)
at $z>4$. Whenever possible, we have obtained near-simultaneous 
optical imaging and spectroscopy with the 8-m Hobby-Eberly Telescope
(Ramsey et~al. 1998), so that the X-ray and optical properties can be 
compared without concerns about variability. 


Seven $z>4$ AGN have been selected via their X-ray emission, spanning the 
redshift range $z=$~4.14--5.19 (Henry et~al. 1994; Zickgraf et~al. 1997; 
Schneider et~al. 1998; Barger et~al. 2002; Silverman et~al. 2002; 
Castander et~al. 2003). These objects suffer from less 
selection bias than AGN found at other 
wavelengths; the $\approx$~2--40~keV \hbox{X-rays} detected from $z>4$ AGN are 
extremely penetrating and thus unlikely to be absorbed. In addition, the 
$z>4$ X-ray selected AGN found in deep X-ray surveys (i.e., the \chandra\
Deep Field-North and the Lockman Hole) are substantially less luminous 
(with $M_{\rm B}\approx -21.4$ to $-24.4$) than those found by the SDSS 
and DPOSS. These moderate-luminosity AGN are likely to be the majority 
AGN population at $z>4$; as such, they deserve at least as much attention 
as the highly luminous SDSS and DPOSS quasars. 


\subsection*{Basic X-ray Properties: Fluxes and Luminosities}

Figure~1b shows the X-ray and optical fluxes for $z>4$ AGN with sensitive
X-ray observations. Most of these objects, including even the optically
brightest ones known, are faint X-ray sources with 0.5--2~keV fluxes of 
$\simlt 3\times 10^{-14}$~erg~cm$^{-2}$~s$^{-1}$. The only exception
to this rule is a small group of radio-loud blazars which have
0.5--2~keV fluxes up to $\approx 6\times 10^{-13}$~erg~cm$^{-2}$~s$^{-1}$
(e.g., Moran and Helfand 1997; Yuan et~al. 2000; Fabian et~al. 2001a,b). 
It is also clear from Figure~1b that, as expected from low-redshift
AGN, there is a correlation between X-ray and optical flux; 
Vignali et~al. (2003a, 2003c) find this correlation to be significant at the
$>99$\% confidence level and give the best-fit relationship (see 
their \S3.1). The scatter around the best-fit relationship, however, 
is substantial (a factor of $\approx 3$ in X-ray flux).
Practically, this scatter makes it risky to attempt long X-ray 
spectroscopic observations before an X-ray flux has been established
via a shorter observation. 

The 2--10~keV rest-frame luminosities of the AGN shown in Figure~1b
span the wide range $\approx 2\times 10^{43}$~erg~s$^{-1}$ (e.g., Vignali et~al. 2002) 
to $\approx 2\times 10^{47}$~erg~s$^{-1}$ (e.g., Fabian et~al. 1998).  
Only the radio-loud blazars inhabit the highest X-ray luminosity range; 
RQQs generally have 2--10~keV X-ray luminosities of 
$\simlt 3\times 10^{45}$~erg~s$^{-1}$ (e.g., Vignali et~al. 2003a).  
These luminosities assume isotropic emission and no amplification
due to gravitational lensing (e.g., Wyithe and Loeb 2002). 

\subsection*{X-ray Contribution to the Spectral Energy Distribution}

In AGN studies the contribution of X-rays to the spectral energy distribution 
is often parameterized with \aox, the slope of a nominal power law between
rest-frame 2500~\AA\ and 2~keV  
[$\alpha_{\rm ox}=0.384\log (f_{\rm 2~keV}/f_{2500~\mbox{\footnotesize \AA}}$) 
where $f_{\rm 2~keV}$ is the flux density at 2~keV and
$f_{2500~\mbox{\footnotesize \AA}}$ is the flux 
density at 2500~\AA].\footnote{For $z>4$ AGN we have 
generally calculated $f_{\rm 2~keV}$ from the observed-frame 0.5--2~keV
flux (assuming an X-ray power-law photon index of $\Gamma=2$).
Thus, the derived \aox\ values are actually based on the relative
amount of X-ray flux in the 0.5$(1+z)$~keV to 2$(1+z)$~keV rest-frame band.}
AGN with larger negative \aox\ values have relatively weaker X-ray emission. 
The optically selected RQQs at $z>4$ have more 
negative \aox\ values ($\langle \alpha_{\rm ox} \rangle=-1.74\pm 0.02$) 
than low-redshift comparison samples 
(e.g., the Bright Quasar Survey RQQs have $\langle \alpha_{\rm ox} \rangle=-1.56\pm 0.02$; 
Brandt et~al. 2000; see Figure~2a). However, the $z>4$ RQQs are 
also more luminous on average (at 2500~\AA; see Figure~2b), and
thus one must address if the difference in \aox\ is due to 
redshift or luminosity (or both). Previous studies of this topic have
generally concluded that \aox\ primarily depends upon luminosity 
rather than redshift (e.g., Avni et~al. 1995 and 
references therein), although Bechtold et~al. (2003) recently 
argued that \aox\ depends primarily upon redshift. Most work
on this topic has been limited by heterogeneous AGN selection
criteria or limited coverage of redshift or luminosity. 

We have been investigating the \aox\ versus redshift/luminosity
issue with a sample of 137 SDSS RQQs at $z=$~0.16--6.28 that
lie in pointed \rosat, \chandra, and \xmm\ observations
(Vignali et~al. 2003b, hereafter VBS; also see Vignali et~al. 2003c). We 
have focused on RQQs since they comprise the bulk ($\approx$~85--90\%; e.g., 
Stern et~al. 2000) of the quasar population and also are ``simpler'' 
systems to study (since it is not necessary to decompose the X-ray 
emission into accretion-disk-linked and jet-linked components).
Compared to much of the previous work, our analysis has several 
significant advantages: 

\begin{enumerate}

\item
The SDSS quasars utilized have been optically selected in a well-defined
manner and span large ranges of redshift and luminosity
(e.g., Schneider et~al. 2002 and references therein).  

\item
Sensitive radio data from the FIRST and NVSS surveys allow effective 
selection of only RQQs. 


\item
We have removed $z\simgt 1.5$ Broad Absorption Line quasars (BALQs)  
and quantitatively constrained the effects that unidentified $z\simlt 1.5$ 
BALQs can have upon our results. The underlying \hbox{X-ray}
emission properties of BALQs are often difficult to measure because
their X-ray spectra are ``corrupted'' by heavy absorption
(e.g., Green et~al. 2001; Gallagher et~al. 2002; see the open squares
in Figure~2). 

\item
We have examined the possible effects of flux amplification due to 
gravitational lensing of $z>4$ quasars (e.g., Wyithe and Loeb 2002).
\aox\ will probably not be changed by gravitational lensing, but the 
true luminosity will differ from that assumed. 

\item
We have adopted the observationally favored ``$\Lambda$CDM'' cosmology 
and have assessed the sensitivity of our main results to the adopted
cosmology. 

\end{enumerate}

\noindent
Partial correlation analyses, using the method of Akritas and Siebert (1996), 
find an anti-correlation between \aox\ and 2500~\AA\ luminosity at the 
3.4--3.9$\sigma$ level (controlling for the effect of redshift). 
By contrast, no highly significant redshift dependence of \aox\ is
found (1.7--2.3$\sigma$; controlling for the effect of luminosity). Thus, 
VBS find results that are qualitatively consistent with those of most
previous workers: \aox\ appears to depend primarily upon luminosity 
rather than redshift.\footnote{We note, however, that there is 
substantial scatter in \aox\ ($\approx \pm 0.2$) at all luminosities 
and redshifts; the correlations are far from being ``tight.''} 

\begin{figure}
\includegraphics[width=90mm,angle=0]{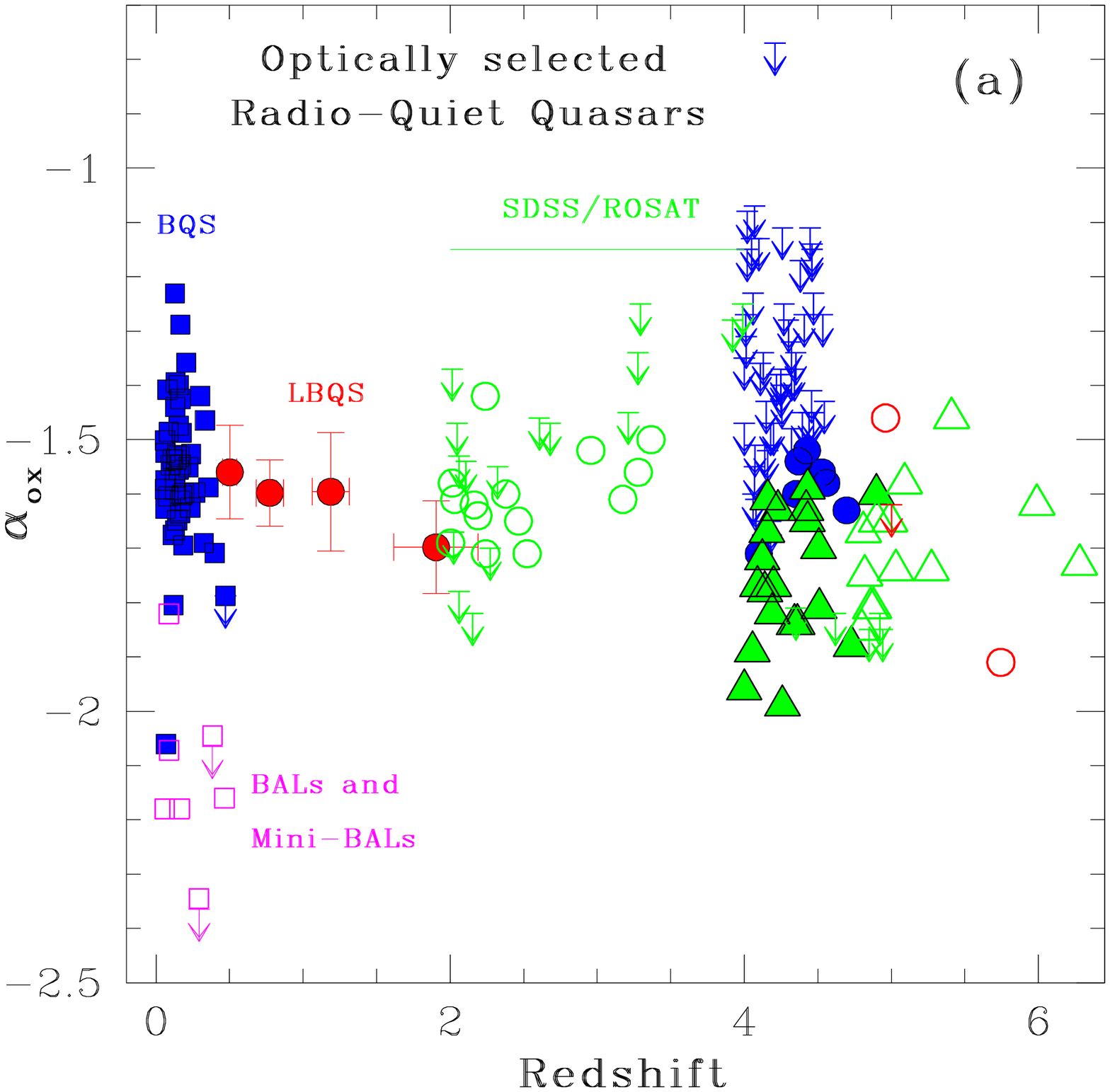}
\includegraphics[width=90mm,angle=0]{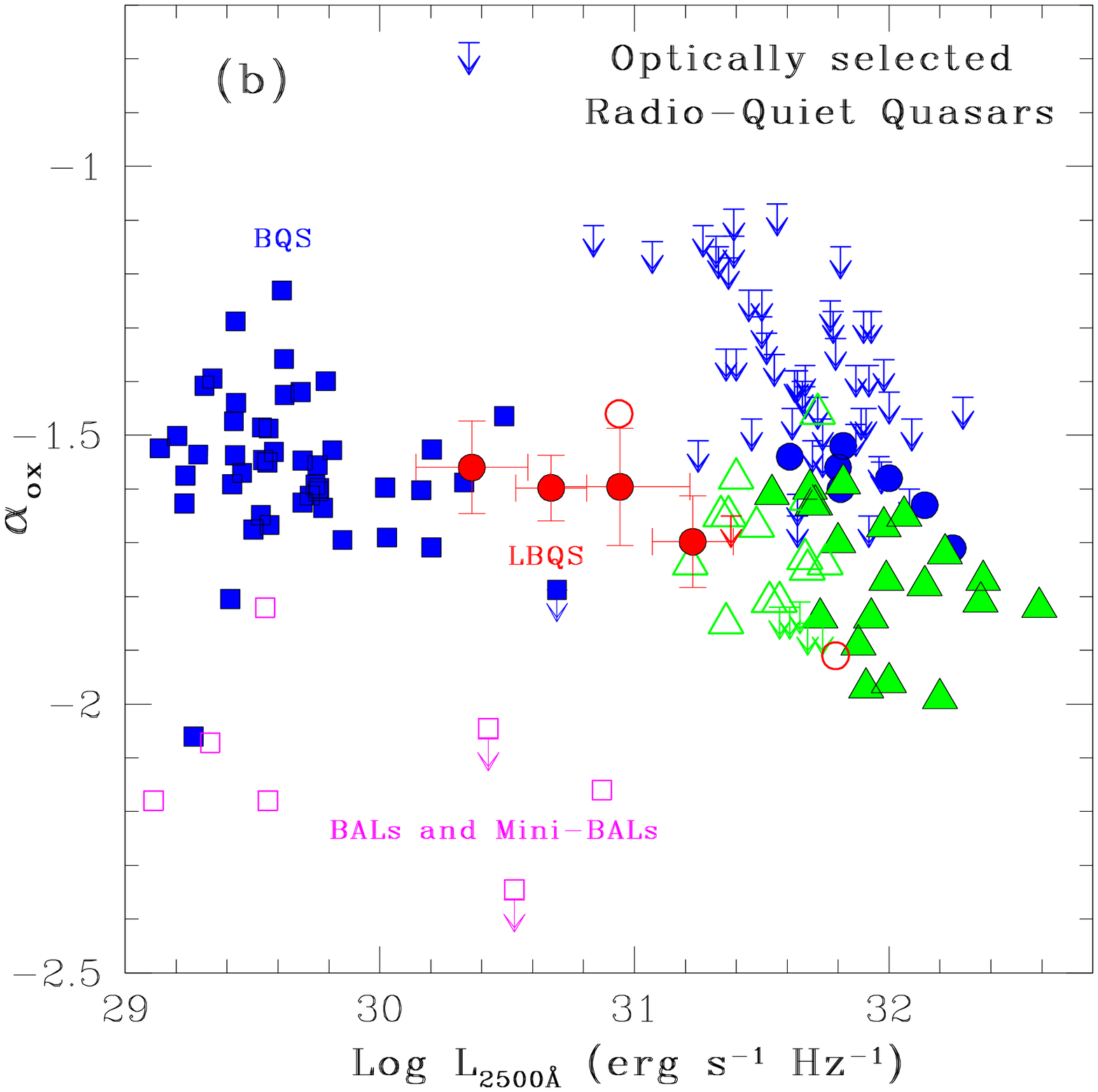}
\vspace*{-0.4 in}
\caption{The parameter \aox\ versus (a) redshift and (b) 2500~\AA\ 
luminosity for optically selected RQQs. The symbols at $z>4$ are 
the same as for Figure~1b. The open squares are for seven luminous, 
absorbed Bright Quasar Survey (BQS) RQQs (BALQs and mini-BALQs), and 
the solid squares are for the other 46 luminous BQS RQQs (from 
Brandt et~al. 2000). The solid circles with error bars show 
stacking results for Large Bright Quasar Survey (LBQS) RQQs from 
Figure~6d of Green et~al. (1995). The data points in (a) from
$z=$~2--4 are taken from the \rosat\ study of SDSS RQQs by
Vignali et~al. (2003b).}
\end{figure}

For $z>4$ AGN, the lack of any strong changes in \aox\ with redshift
(after controlling for luminosity effects) is generally consistent 
with the lack of strong spectral evolution at other wavelengths. For 
example, the ultraviolet rest-frame spectra of $z>4$ quasars are only
subtly different from those at lower redshifts 
(e.g., Schneider et~al. 1989; Constantin et~al. 2002), 
and the fraction of RLQs does not appear to change significantly with 
redshift (e.g., Stern et~al. 2000). 

\subsection*{X-ray Spectroscopy}

\begin{figure}
\includegraphics[width=80mm,angle=-90]{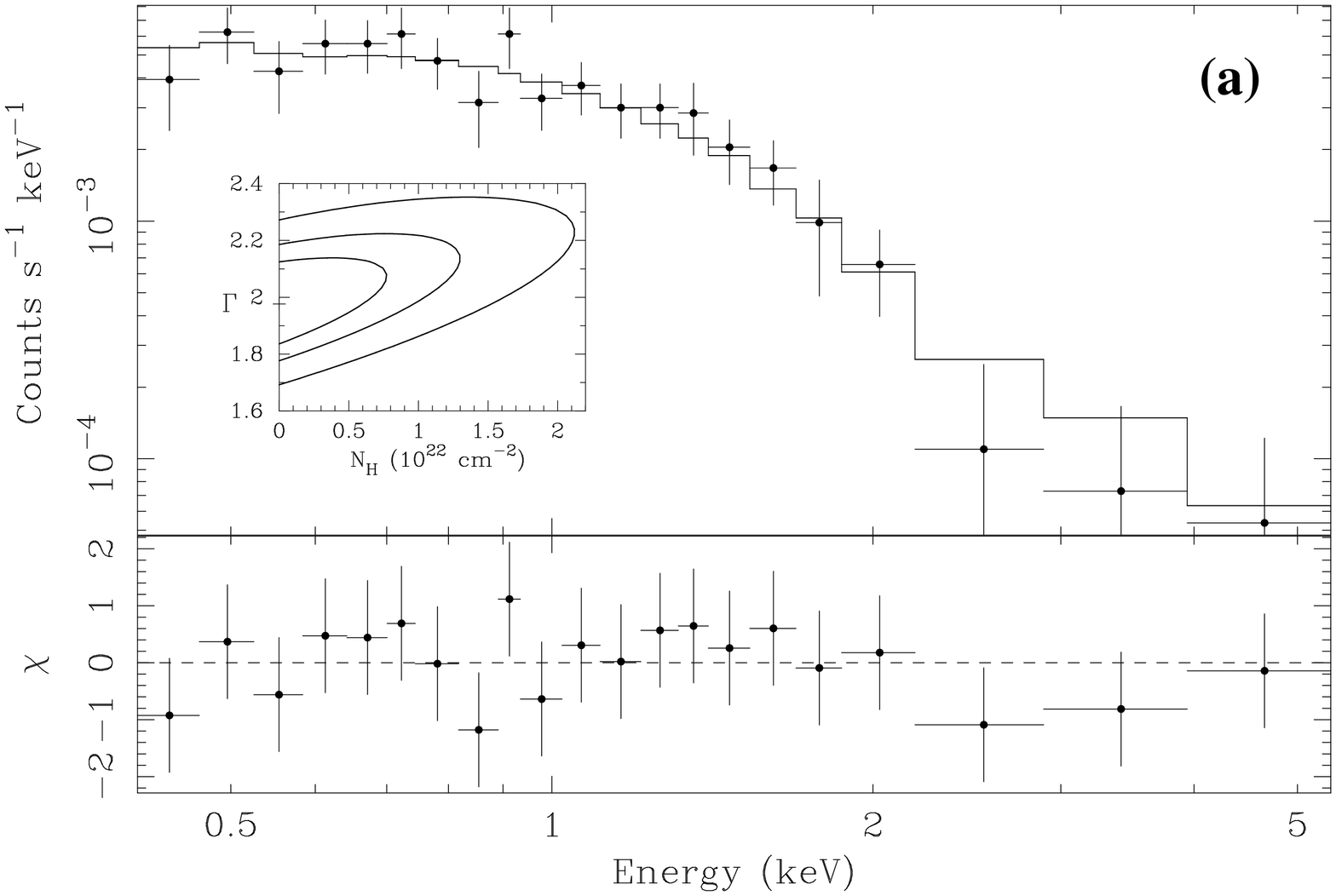}
\hfill
\includegraphics[width=70mm,angle=-90]{brandt-fig03b.ps}
\vspace*{-0.6in}
\caption{(a) Stacked X-ray spectrum constructed from nine luminous
DPOSS quasars at $z=$~4.09--4.51 fitted with a power-law model
and Galactic absorption (see the text). Data-to-model residuals,
in units of $\sigma$, are shown in the bottom panel. The 
insert shows the 68, 90, and 99\% confidence regions for photon
index and intrinsic column density obtained via fitting
with the Cash (1979) statistic. 
(b) Hard X-ray photon index versus redshift. The solid circles
are the 2--10~keV data points with the smallest statistical errors
from Page et~al. (2002). The stars are, from left to right, 
the data points from
APM~08279+5255 at $z=3.91$ (Chartas et~al. 2002), 
nine luminous DPOSS quasars at $z=$~4.09--4.51 (Vignali et~al. 2003a; 
plotted at their average redshift), and 
CXOHDFN~J123647.9+620941 at $z=5.186$ (Vignali et~al. 2002).}
\end{figure}

X-ray spectral analyses of $z>4$ AGN can determine the photon index
of the X-ray power law and the properties (column density and ionization
level) of any intrinsic absorption. Aside from the small group of radio-loud 
blazars mentioned above, detailed X-ray spectral analyses of individual 
$z>4$ AGN have not yet been performed due to their low X-ray fluxes 
(although see Chartas et~al. 2002 and Hasinger et~al. 2002
for X-ray spectral analyses of the $z=3.91$ gravitationally lensed 
BALQ APM~08279+5255).  

It has been possible, however, to derive average X-ray spectral 
constraints via the joint fitting of several $z>4$ AGN (using the 
Cash 1979 statistic). Such joint fitting can provide a more stable 
estimate of average X-ray spectral properties than 
a single-object spectrum of the same statistical quality (since any 
single object might be atypical). Vignali et~al. (2003a) have recently
performed joint \chandra\ spectral fitting of nine luminous DPOSS quasars 
at $z=$~4.09--4.51; together these quasars have $\approx 340$ counts, and 
only $\approx 2$ background counts are expected. The results are shown 
in Figure~3a and constrain the rest-frame 2--30~keV emission. The spectra
are well fit by a power-law model with only Galactic absorption. 
The derived power-law photon index is $\Gamma=1.98\pm 0.16$ (errors
are for 90\% confidence), entirely consistent with those found for
$z\approx$~0--3 samples of RQQs in the 2--10~keV band (see Figure~3b; 
e.g., George et~al. 2000; Reeves and Turner 2000; 
Page et~al. 2002).\footnote{We note that, at present, there is no
evidence that the photon index of the underlying X-ray 
power law has a luminosity dependence (after allowing for
Compton-reflection effects).} 
The average constraint on neutral intrinsic absorption, assuming 
solar abundances, is $N_{\rm H}<8.8\times 10^{21}$~cm$^{-2}$. If
X-ray absorption were present at the level found in high-redshift 
RLQs, it should have been detected. 
Vignali et~al. (2003c) have also used joint \chandra\ spectral 
fitting to constrain the average spectral properties of 13 SDSS quasars 
at $z=$~4.81--6.28. These objects are well fit by a $\Gamma=1.84\pm 0.31$
power law and Galactic absorption

We note that Bechtold et~al. (2003) have reported X-ray spectral
constraints for 16 luminous quasars at $z=$~3.67--6.28. They do not
report the results of a joint spectral analysis, but the mean of the 
photon indices in their Table~2 is $\langle\Gamma\rangle=1.50\pm 0.15$. 
This mean photon index is significantly harder 
(at the $\approx 2.5\sigma$ level) than those reported
above for the DPOSS and SDSS quasars.


Vignali et~al. (2002) have performed X-ray spectral analyses of
the three $z>4$ moderate-luminosity AGN found in the \chandra\ Deep 
Field-North; such analyses are only possible due to the extremely
deep 2~Ms exposure. 
The highest quality X-ray spectrum was obtained for the low-luminosity
quasar CXOHDFN~J123647.9+620941 ($z=5.186$; $M_{\rm B}=-23.4$; 138 counts) 
and is well fit by a $\Gamma=1.81\pm 0.31$ power-law model with Galactic 
absorption. 
The other two AGN, 
CXOHDFN~J123642.0+621331 ($z=4.424$; $M_{\rm B}=-21.6$; 45 counts) and
CXOHDFN~J123719.0+621026 ($z=4.137$; $M_{\rm B}=-21.4$; 117 counts), 
have Seyfert-like luminosities. They have flatter X-ray spectra with 
effective photon indices of $\Gamma\approx$~1.6 and $\Gamma\approx$~1.1,
respectively, when only Galactic absorption is assumed in the fitting
(these fits are statistically acceptable). The X-ray spectrum of 
CXOHDFN~J123719.0+621026 can also be acceptably modeled with a 
$\Gamma\approx$~2.0 power law and an intrinsic column density of
$\approx 2\times 10^{23}$~cm$^{-2}$. 

\subsection*{X-ray Variability}

At present, information on the X-ray variability of $z>4$ AGN is very limited. 
The snapshot X-ray observations used to detect most $z>4$ quasars to date (see above)
cover timescales of only $\sim 20$~minutes in the rest frame, and most luminous
quasars at low redshift do not vary strongly on such short timescales. Highly 
significant X-ray variability at $z>4$ has only been detected from the blazars
PMN~J0525--3343 (Fabian et~al. 2001b), 
GB~1428+4217 (Fabian et~al. 1999; Boller et~al. 2000), and 
GB~1508+5714 (Moran and Helfand 1997). 
CXOHDFN~J123719.0+621026 in the \chandra\ Deep Field-North may also be
X-ray variable (Vignali et~al. 2002). 
There is a tentative claim that quasar X-ray variability increases 
with redshift in the sense that quasars of the same X-ray luminosity
are more variable at $z>2$ (Manners et~al. 2002).


\subsection*{X-ray Imaging}

X-ray imaging of $z>4$ AGN can allow searches for spatial extent due to
gravitational lensing, small-scale X-ray jets, or X-ray scattering
by dust in the intergalactic medium. 
The only $z>4$ AGN with published 
evidence for X-ray spatial extent is SDSS~J103027.10+052455.0 at $z=6.28$,
and Schwartz (2002) suggests that this extent may be due to gravitational
lensing. The putative X-ray extent needs to be confirmed with a deeper 
\chandra\ observation, and lensing will be tightly constrained with an
\hst\ observation. 
Telis et~al. (2003) have set a direct upper limit on the density of
cosmological dust from the absence of an X-ray scattering halo around
the $z=4.30$ blazar GB~1508+5714; they find that intergalactic 
extinction is unlikely to have significantly affected the Hubble 
diagram of Type~1a supernovae. 

X-ray imaging can also search for companion objects such as clustered
AGN in large-scale structures (e.g., Djorgovski 1999; Djorgovski et~al. 2003) 
or large-scale X-ray jets. Schwartz (2002) tentatively 
reported the detection of a large-scale X-ray jet associated 
with the $z=5.99$ quasar SDSS~J130608.26+035626.3, although such a
detection would have been somewhat surprising given that this object (and
its putative jet) is not a powerful radio source (i.e., its 
Kellermann et~al. 1989 radio-loudness parameter is $R<8.7$). 
Deep \hbox{$I$-band} imaging of the putative jet (Ivanov 2002) suggests the
X-ray emission is probably associated with an unrelated foreground 
galaxy. Nevertheless, jets should be searched for from other $z>4$ AGN. 
Vignali et~al. (2003a) and Vignali et~al. (2003c) have searched 
for statistical overdensities of companion X-ray sources near DPOSS and 
SDSS quasars, but no overdensities are found. 



\section*{CONCLUSIONS AND FUTURE PROSPECTS}

\

From a general perspective, the main conclusions from the X-ray analyses
reviewed above are the following: 

\begin{enumerate}

\item
To the limits of current observation, AGN at $z\approx$~4--6 and
$z\approx$~0--3 have reasonably similar X-ray and broad-band spectra
(after controlling for luminosity effects upon the broad-band spectra). 
Specifically, the distributions of X-ray power-law photon index 
and \aox\ appear consistent. 

\item
The current X-ray data do not provide any hints for different
accretion mechanisms at low and high redshift. The X-ray data are 
consistent with the idea that early black holes at high redshift feed 
and grow in the same way as local ones. If the accretion mechanism
does change at high redshift, it does so without dramatically 
affecting the X-ray emission made in the immediate vicinity of
the black hole. 

\item
The small-scale X-ray emission regions of AGN appear relatively
insensitive to the strong large-scale environmental changes that
have occurred from $z\approx$~0--6. The quasar population evolves
strongly over this time, but individual quasar X-ray emission 
regions appear to evolve much less. 

\item
The X-ray data provide support that, at least to first order,
the spectral energy distributions of local AGN can be used when 
(1) estimating the black hole masses of high-redshift AGN 
via the Eddington argument [$\approx$~(1--5)$\times 10^9$~M$_\odot$ 
for luminous SDSS quasars; e.g., Fan et~al. 2001] and 
(2) calculating the effects of early AGN upon the intergalactic medium 
(e.g., Venkatesan et~al. 2001; Cen 2003; Machacek et~al. 2003).

\item
X-ray emission appears to be a universal property of luminous
AGN, even at $z\approx$~4--6. The X-rays also do not appear 
to be generally blocked from view by material in nascent AGN host 
galaxies.\footnote{Due to its low metallicity and relatively low 
integrated column density, the intergalactic medium should not 
prevent the X-ray detection of AGN even to very high redshift 
(e.g., Aldcroft et~al. 1994; Miralda-Escud\'e 2000).} 

\end{enumerate}

\noindent
There are a number of ways the work described above can be extended
including the following: 

\begin{enumerate}

\item
At present, there is only a limited number of X-ray detections at $z>5$
(see Figure~1a). It is important to increase this number as further $z>5$
AGN are discovered, so that strong statistical conclusions can be drawn 
about X-ray emission at the highest 
redshifts. For example, the SDSS should find $\approx 100$
quasars with $z>5$ and $\approx 30$ with $z>6$; it can find quasars
up to $z\approx 6.5$ (e.g., Fan et~al. 2003). 
Hopefully X-ray (e.g., Alexander et~al. 2001; Barger et~al. 2003)
and infrared (e.g., Warren 2002) selection 
can be exploited to discover even higher redshift AGN. Ultimately, 
with missions such as \genx, it should be possible to detect 
$\sim 10^4$~M$_\odot$ black holes in proto-AGN at $z\sim 10$; these 
objects will have X-ray fluxes 
of $\sim 5\times 10^{-20}$~erg~cm$^{-2}$~s$^{-1}$  
(compare with Figure~1b).\footnote{For further discussion
see the \genx\ white paper submitted to the NASA Structure 
and Evolution of the Universe roadmap at 
http://universe.gsfc.nasa.gov/docs/roadmap/submissions.html.}

\item
Minority AGN populations at $z>4$ need to be investigated
better in the X-ray regime. These include 
quasars lacking emission lines (e.g., Fan et~al. 1999; Anderson et~al. 2001; Vignali et~al. 2001), 
BALQs (e.g., Brandt et~al. 2001; Goodrich et~al. 2001; Vignali et~al. 2001, 2003a, 2003c), and
RLQs. 


\item
The constraints on \aox\ evolution with redshift and luminosity can 
be greatly improved as the SDSS AGN sample increases in size. Ultimately 
it should be possible to perform analyses like those in VBS with 
$\sim 3,400$ SDSS RQQs. 
Further X-ray studies of faint, moderate-luminosity AGN at $z>4$ 
will also be very useful for breaking the degeneracy between redshift
and luminosity. 

\item
The current X-ray spectral constraints on $z>4$ AGN are admittedly limited;
these need to be improved. In addition to the underlying power law, 
low-redshift AGN show X-ray spectral complexity including 
neutral and ionized absorption, 
iron~K$\alpha$ lines,
reflection continua, and
soft excesses. 
The \hbox{X-ray} constraints on neutral absorption are already becoming 
interesting, but constraints on other types of spectral complexity
are weak. 
With \hbox{40--100~ks} exposures, \xmm\ can obtain moderate-quality 
($\approx$~1000--2000 count) $\approx$~1--50~keV rest-frame spectra 
of luminous $z>4$ quasars with 0.5--2~keV fluxes of 
$\simgt 1.5\times 10^{-14}$~erg~cm$^{-2}$~s$^{-1}$ (see Figure~1b).
Efficient X-ray spectroscopy of the majority of the $z>4$ AGN
population, however, will require future missions such as \conx, 
\xeus, and \genx\ (see Vignali et~al. 2001, 2003a for relevant 
simulations).  

\item
Systematic variability studies of $z>4$ AGN would allow the tentative 
claim that quasar X-ray variability increases with redshift 
(Manners et~al. 2002) to be tested in extremis. 

\item
Further X-ray imaging studies of $z>4$ AGN will ultimately help to pin down
the incidence of \hbox{X-ray} jets and gravitational lensing. Constraints will
also be placed upon the amount of dust in the intergalactic medium and
AGN clustering in large-scale structures. 

\end{enumerate}


\section*{ACKNOWLEDGMENTS}

\

We thank all of our collaborators on the work reviewed here. 
We thank 
G. Brunetti,
S.C. Gallagher,  
D.A. Schwartz, and 
M.A. Strauss for useful discussions. 
We gratefully acknowledge the financial support of 
NASA LTSA grant NAG5-8107 (WNB, CV), 
NASA grant NAG5-9918 (WNB, CV, DPS), 
\chandra\ X-ray Center grant G02-3134X (WNB, CV, DPS), 
NSF grant AST99-00703 (DPS),
\chandra\ X-ray Center grant G02-3187A (DMA, FEB), and 
NASA grant NAS~8-01128 (GPG). 



\end{document}